# Architecture for Analysis of Streaming Data

Sheik Hoque* and Andriy Miranskyy†
*† Department of Computer Science, Ryerson University, Toronto, Canada
* Royal Bank of Canada, Toronto, Canada
* sheik.hoque@ryerson.ca, † avm@ryerson.ca

*Abstract*—While several attempts have been made to construct a scalable and flexible architecture for analysis of streaming data, no general model to tackle this task exists. Thus, our goal is to build a scalable and maintainable architecture for performing analytics on streaming data.

To reach this goal, we introduce a 7-layered architecture consisting of microservices and publish-subscribe software. Our study shows that this architecture yields a good balance between scalability and maintainability due to high cohesion and low coupling of the solution, as well as asynchronous communication between the layers.

This architecture can help practitioners to improve their analytic solutions. It is also of interest to academics, as it is a building block for a general architecture for processing streaming data.

## I. INTRODUCTION

Streaming analytics is playing an important role in solving problems in various domains, e.g., monitoring climate, fraud detection, and health management (see [12] for details). The streaming analytics market is growing rapidly: from $3.08 billion in 2016 to $13.7 billion by 2021 (estimated) [1].

Data from a stream could be interpreted differently by different use cases. For example, in the case of an e-commerce website, submitting an order can be a trigger to 1) a warehouse system to update inventory, 2) a user recommendation system to provide suggestions to like-minded users, and 3) a customer profiling system to provide additional recommendations of similar products.

Not every use case is interested in all the incoming data; rather they may be looking for a particular event to occur. For example, for the e-commerce website use case, an analytic model interested in sales of grocery products has no interest in data on sales of electronics.

Therefore, it is sub-optimal to feed all the data to all the analytic models. Only a relevant subset of data should be delivered to a program dealing with a particular use case. In this paper, our **goal** is to introduce a layered architecture that 1) takes heterogeneous streaming data from various sources as input, 2) identifies a subset of these data relevant to our business use cases, and 3) performs analysis of the data to satisfy these use cases.

The architecture consists of a combination of microservices [13] and instances of the publish-subscribe pattern [6], spread over multiple layers.

A microservice encapsulates a set of tightly coupled features of an application. It can serve the functionality of these features as a single entity. An architecture consisting of microservices decentralises and separates loosely coupled features and thus avails the parallelisation of development as well as testing [13].

A publish-subscribe pattern is an asynchronous communication hub [6]. An instance of this pattern transmits messages from publishers to subscribers. It makes the communication decoupled and scalable, as a publisher does not require to keep a roster of its consumers. Moreover, a subscriber is not required to poll a publisher periodically hence the reduction of response time and delivery latency.

An architecture that has these two building blocks yields a good balance between maintainability and scalability, as will be discussed below. We also need to specify the number of layers in the architecture. Our exploration of multi-layered architectures shows that the 7-layered one gives good cohesion and coupling.

The rest of the paper is organised as follows. Section II defines various architectures, while Section III applies 7-layered architecture to a business scenario. Section IV provides related literature. Finally, Section V concludes the paper.

## II. ARCHITECTURES

Business use cases dealing with analysis may vary, ranging from sentiment analysis of a product, to supply chain analysis, to prediction of a future stock price. Thus, the output from the models will vary, ranging from business intelligence report (e.g., identifying potential cases to improve a business), to automated notification to a warehouse (e.g., telling when to replenish the stock), to an order to a trading platform (e.g., directing to buy or sell a stock).

**Example II.1.** Suppose an analyst would like to estimate stock price of a company. One can aggregate Rich Site Summary (RSS) news feeds targeting global economic analysis, political feeds, newspaper and magazine articles about the economic sector in which the company operates, news feed about company itself, etc. All of the information will then be fed into an analytic model that will try to predict the movement of stock price, based on historical information. We are going to use this example throughout the paper.

This is a relatively straightforward problem, when dealing with a single financial product or stock. However, a large organisation typically needs to perform, in parallel, analysis of multiple products and/or stocks. This implies that the same piece of data may be used by multiple models. For example, if we are to estimate stock price of multiple companies, then news feeds related to global economic and political analysis would be relevant to each of these companies.

One can build a solution involving multiple models as a monolith application. However, this will lead to a number of challenges. For example, scalability will be poor (as vertical scaling is expensive, while horizontal scaling is fragile [23]); maintainability and comprehensibility will be low finally, reliability and availability will be inadequate. Thus, one needs to reduce the coupling of the components that implement various features of the analytic system. There are multiple ways to reach this goal.

For example, one can start separate processes for different tasks, making monolithic system multi-processed (and, potentially, multi-threaded). This solution scales vertically but has challenges with horizontal scalability (as it is challenging to reliably deploy processes on multiple compute nodes).

Another example is using microservice-based architecture, which helps to overcome horizontal scalability challenge and improve maintainability. However, tightly coupled microservices (with intense flow of data between them) impact solution's performance. Moreover, communication dependency risks the maintainability of the system (as the break of communication may lead to loss of data). Thus, microservices need to be loosely coupled, and a communication failure between a pair of microservices of a system should not bring the full system down. An asynchronous communication[1] is a good option to avoid such a failure.

Thus, "sharding" the functionality using microservices and passing information between these microservices, using an instance of publish-subscribe pattern, yields a good balance between scalability, maintainability, comprehensibility, reliability, and availability.

Assuming that our architecture is layered, how many layers do we need and which functionality goes into which layer? Below we will explore a number of cases.

As mentioned in Section I, the building blocks of our architecture are publish-subscribe pattern and microservices. The primary focuses of the design are 1) to move only relevant data towards a microservice based on its use case(s) and 2) to make the components independent of each other and hence easily pluggable. A combination of both of these has a good impact on performance of the system with high throughput and low latency. On top of that, pluggable feature eases the deployment, maintenance, and regression testing efforts.

Formally, the number of layers in our architecture can be summarised using the formula $2n+1$. Odd layers perform data transformation, while even layers serve as a communication mechanism. That is, $i$-th layer will pass messages from $i-1$ layer to $i+1$ layer. We consider cases $n = 0, 1, \ldots, 4$ below. To preserve space, we give detailed description only of the $n = 3$ case and a cursory one of the remaining cases.

---

[1]Note that asynchronous communication would also help to improve scalability of a monolith application, but the issue of maintainability/comprehensibility would remain.

### A. 1-layered architecture ($n = 0$)

This case yields one layer. It is a special case, where we do not have a separate communication layer. For every analytic model, we will have an individual microservice; data extraction, transformation, filtering, and analysis for a given model will be implemented in the same microservice.

This solution is better than monolith; the pros are as follows. We can scale each microservice individually. The code is easier to maintain and comprehend (as the source code for each use case is maintained separately). The solution is more reliable and available: if we break one microservice, the rest will still be functioning.

There are also cons. This solution leads to computational overhead, as all the programs will have to duplicate their data extraction efforts. Each program will have to monitor every data input to make sure that all the relevant data points (e.g., news articles relevant to a given model) are captured.

### B. 3- and 5-layered architectures ($n = \{1, 2\}$)

As we increase the number of layers, we will continue separating the functionality to reduce coupling and to increase cohesion of the components of our solution.

In the case of the 3-layered architecture, the first layer extracts the data from various input feeds and publishes the data to topics of the second layer, the third layer contains microservices implementing analytic models. They listen to the relevant topics in the second layer and execute analytic models, when new data get published into a topic of interest.

In the case of the 5-layered architecture, the first four layers further decouple data extraction and transformation logic. Layers 2 and 4 implement publish-subscribe pattern. The first layer listens to inputs, transforms data to universal format, and publishes the data to the topics maintained by the second layer. The third layer listens to the topics of the second layer, applies business logic on a received message, categorises the message, and publishes it to a topic maintained in the fourth layer. The last (fifth) layer consumes the data from the relevant topics in the fourth layer and executes the model.

In addition to pros that 1-layered architecture has, the 3- and 5-layered architecture split data extraction and transformation logic from the analysis one. This further increases cohesion and decreases coupling of the layers, improves code comprehensions (as the amount of code per component decreases) and maintainability.

However, cohesion and coupling are not ideal yet, as the model's analytic and data gathering logic are still bundled together in the last layer, hence the need for an additional pair of layers.

### C. 7-layered architecture ($n = 3$)

To further improve coupling and cohesion, we factor out (into separate layers) logic related to core data extraction, transformation, and analysis activities, namely, converting, splitting, aggregating, and modelling. The layers communicate with each other with an instance of publish-subscribe pattern.

A diagram of the 7-layered architecture is given in Fig. 1. Details are provided below.

*1) Converter:* The first layer listens to $L$ data streams/feeds, converts the data to universal format, and publishes the data to topics maintained by the second layer. The second layer implements publish-subscribe pattern and hosts $L$ topics, working as a data transmission channel between the first and the third layer. There is a 1-to-1 relation between microservices in layer 1 and topics in layer 2.

An example of conversion logic is given below. A system can have multiple input sources, such as RSS feeds, Twitter feeds, and files in different formats (e.g., PSV, Parquet, JSON, and Avro). These different formats will be transformed to a universal format (e.g., JSON) and then resulting data will be published to a topic of the second layer. The benefit of having a universal format is that downstream microservices and publish-subscribe layers will be able to interact in the same format.

*2) Splitter:* $L$ microservices in the third layer listen to $L$ topics of the second layer. There is a 1-to-1 mapping between topics of the second layer and microservices of the third layer. Each microservice applies business logic on a received message, categorises the message, and publishes it to one or more topics of the fourth layer. Those messages that do not relate to any topic are discarded. Thus, there is a 1-to-many relation between microservices in the third layer and topics in the fourth layer.

The fourth layer implements publish-subscribe pattern, hosting $X$ topics, and works as a data transmission channel between the third and the fifth layers. The actual value of $X$ is typically independent of the number of input streams and models and is dictated by the nature of use cases and the data.

Going back to Ex. II.1: a company can be categorised based on the sector where it operates (e.g., energy, healthcare, or technology). Suppose our analytic models, predicting stock prices of companies, are interested in the news articles only for the sector in which a given company operates. Then, to reduce the amount of computations by each model (which listens to the news articles coming to various topic), a microservice in the third layer will take an article from the "news articles" topic in the second layer, identify sectors discussed in this article, and then publish the article to the topics managed by the fourth layer. Say, if an article discusses healthcare and technology sector, then it will be published to two topics: "healthcare" and "technology".

*3) Aggregator:* The logic specifying a list of topics of interest for a given analytic model resides in the fifth layer. Microservices in this layer will listen to one or more topics in the fourth layer, aggregate the messages from multiple topics into one topic per analytic model, and publish the messages to topics in the sixth layer. There is one microservice per analytic model. The total number of models is denoted by $N$. Thus, there will be $N$ microservices in this layer. There is a many-to-many relation between the topics in the fourth layer and the microservices in the fifth layer.

For example, if an analytic model from Ex. II.1 is interested in predicting the stock price of a healthcare company Acme, a microservice implementing data aggregation logic for this model in the fifth layer will listen to topics "healthcare" and "global economy" in the fourth layer, focusing on the news related to Acme[2]. Then the microservice will post a message to a topic in the sixth layer.

The sixth layer implements publish-subscribe pattern, one topic per model, hence $N$ topics, and works as a data transmission channel between layers 5 and 7. There is a 1-to-1 relation between microservices in layer 5 and topics in layer 6, as well as between topics in layer 6 and microservices in layer 7.

*4) Modeller:* The seventh layer implements $N$ analytic models, one microservice per model, consumes the data from the relevant topics in the sixth layer, and executes the model. Note that these microservices may also access external services, e.g., persistent storage[3] that contains older versions of the models, historical data, etc.

*5) Pros vs. Cons:* This architecture has each particular data extraction and transformation feature residing in its own layer. This leads to further increase of cohesion and decrease of coupling of the layers, improves code comprehensions (as the amount of code per microservice decreases) and maintainability in comparison with the 5-layered architecture.

However, the disadvantage of this architecture is data duplication. In the above example, the same news article was published to two topics ("healthcare" and "global economy"). This leads to increased space requirement, as we now need to store (in memory and/or in persistent storage) the same article in two topics. It also leads to increase of the overall number of messages (and transactions per unit of time) that have to be handled by the publish-subscribe software. We will analyse when this trade-off is acceptable from computational perspective in Section II-E.

*D. $n > 3$*

If a solution has a more complex data extraction and transformation logic, one may add additional layers. For example, we can partition Splitter logic into separate layers (say, to separate coarse-grained splitting from fine-grained splitting).

Going back to Ex. II.1, if $n = 4$, "healthcare"-related articles posted by layer 3 into layer 4, will be further split into topics "biotechnology" and "healthcare equipment" by layer 5 into layer 6, before being aggregated in layer 7 (which will post aggregated data into layer 8, which finally will be consumed by models in layer 9). However, we found that $n = 3$ was sufficient for the problems that we faced.

---

[2] Obviously, an article processed by third layer may end up in both topics (say, we may have a long-read on the global economy with some references to the healthcare sector). A microservice in the fifth layer can detect such duplicate artefacts by checking unique id of an article (essentially, one can track the complete chain of transformations associated with a given message using Breadcrumbs concept [2]) and either eliminate or keep a duplicate record (depending on the needs of an analytic model that it is serving).

[3] It can also be used to save new data points and update model's state.

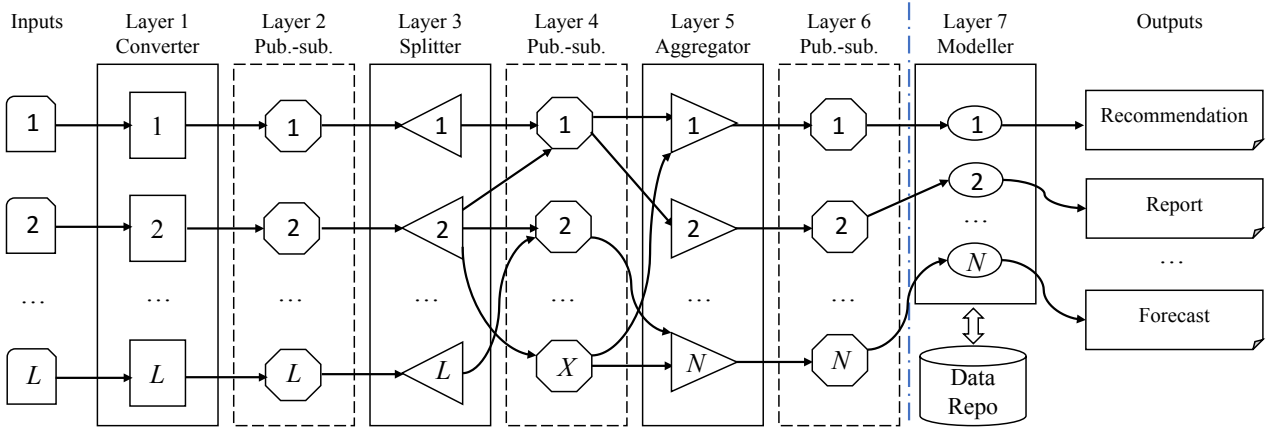

Fig. 1. A diagram of the 7-layered architecture. Dashed lines represent publish-subscribe layers. Vertical dash-dotted line separates data preparation layers from the analytic layer. Arrows denote flow of data. "Data repo" represent persistent storage that the models may access to get or set additional data (as discussed in Section II-C4).

*E. Analysis of complexity*

We performed worst-case scenario analysis, comparing 1- and 7-layered architecture. Details of the analysis can be accesses in the supplementary materials [9].

In nutshell, the order of computational complexity for both architectures is $O(MN)$, where $M$ is the total number of messages in all input feeds. However, the key difference is in the form of the functions governing running time of processing the data in the 1- and 7-layered architectures (denoted by $T^1$ and $T^7$, respectively), hence the difference in performance.

The duplication (as the analysis shows) is driven by the messages published by microservices of layer 3 into topics of layer 4. Let $\gamma$ denote the total number of messages in layer 4. In the worst-case scenario, $\gamma = MX$, i.e., every message is published to every topic.

However, we have never seen such a scenario in practice, as the number of messages that would be relevant to each and every topic of interest is infinitesimally small. Moreover, large number of messages gets discarded, as they have no relevance to the topics of interest, leading to further decrease of $\gamma$. Thus, formally $\gamma = \omega\delta$, where $\omega \in (0, X]$ is the average number of posts per message that do not get discarded, and $\delta \in [0, 1]$ is the fraction of the messages that are not discarded (on average).

We answer the question of when the 7-layered architecture produces results faster than the 1-layered one, by exploring under which conditions $T^1 > T^7$ holds (for the worst-case scenario) in [9]. The analysis shows that the inequality holds when

$$\omega\delta < (cN - c_1 - c_2 - c_3)/(c_5 N + c_6 N + c_4), \quad (1)$$

where $c$ is the highest cost of extracting, transforming and filtering a message in the 1-layered architecture and $c_i$ is the highest cost of performing an operation in the $i$-th layer of the 7-layered architecture. Given that $c$ is the cost of extracting, transforming, and filtering data,

$$c \leq c_1 + c_3 + c_5. \quad (2)$$

This fact, in a limit as $N \to \infty$, simplifies Eq. 1 to

$$\omega\delta < (c_1 + c_3 + c_5)/(c_5 + c_6). \quad (3)$$

This gives us an estimate of the maximum value of $\omega\delta$ when 7-layered architecture will be faster than 1-layered one for a large value of $N$.

For example, if $c_1 = c_3 = c_5 = c_6 = 1$, then $\omega\delta < 1.5$. That is if we retain 80% of the messages ($\delta = 0.8$), and the value of $N$ is large, and the average number of posts to layer 4 for a remaining message is less than 1.9 ($\approx 1.5/0.8$) topics, then 7-layered architecture is more efficient.

Let us explore the relation between $\omega\delta$, $N$, and $c_i$s graphically, by plotting Eq. 1 (where we substitute $c$ with its upper bound from Eq. 2). We plot two cases in Fig. 2: 1) when $c_i = 1$ for all $i$ and 2) when $c_1 = 10$, $c_2 = c_4 = c_6 = 1$, $c_3 = c_5 = 5$. The former setup simulates a case when the cost of all operations is identical, the latter – when the extraction and transformation is the most expensive, assigning messages to topics is cheaper, and passing messages is the cheapest. For a large $N$ (as per Eq. 3), $\omega\delta < 1.5$ and $\omega\delta < 3.3$, respectively.

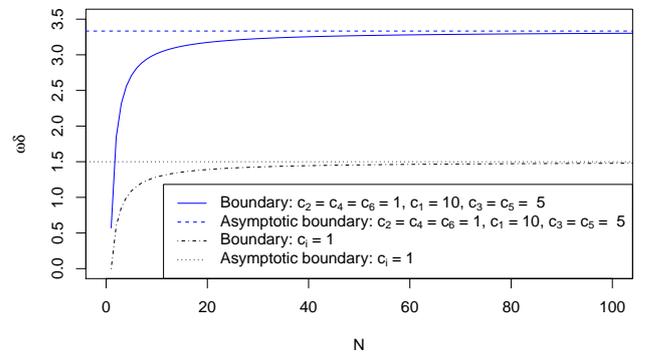

Fig. 2. Plotting $N$ vs. $\omega\delta$ in Eq. 1 for different values of $c_i$. Area under a given curve represents $N$ and $\omega\delta$ values where the 7-layered architecture is faster then the 1-layered one for a given set of $c_i$s. Horizontal lines represent the asymptotic limiting values given by Eq. 3.

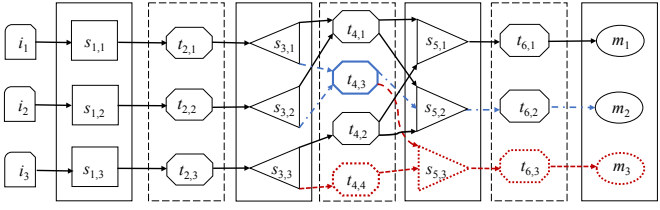

Fig. 3. Architecture for trading application scenario. Objects and relations introduced in v.1 are represented using black solid lines, introduced in v.2 – using blue dash-dotted lines, and in v.3 – using red dotted lines.

Fig. 2 suggests that we reach these limiting values for a relatively small $N$: when $N = 100$, $\omega\delta \approx 1.48$ for the former and $\omega\delta \approx 3.3$ for the latter case. That is Eq. 3 can serve as an upper bound approximation for Eq. 1 for a large $N$.

The actual values of $c_i$s would vary depending on the business use cases; thus, one has to recompute Eq. 1 and/or Eq. 3.

## III. BUSINESS SCENARIO: TRADING APPLICATION

To conserve space, we present below a single scenario based on Ex. II.1, discussing how it can be implemented using the 7-layered architecture. Note that the same principles would apply to scenarios from other industries. An instance of architecture for this example is given in Fig. 3.

*1) Version 1:* We are tasked to build a solution, whose objective is to execute different analytic model predicting prices of financial instruments. Suppose we need to 1) estimate the price of a healthcare company Acme (that we introduced in Section II-C3), which will be done by an analytic model $m_1$ and 2) predict the price of a healthcare sector stock portfolio, which will be performed by an analytic model $m_2$.

Suppose we have three input streams: two RSS news feeds ($i_1$ and $i_2$) from two news agencies and one stock price data feed ($i_3$) from a stock exchange. Then, we will implement three microservices in layer 1 (Converter), listening to each individual input stream. Microservices $s_{1,1}$ and $s_{1,2}$ will receive news articles (from $i_1$ and $i_2$, respectively) and microservice $s_{1,3}$ will get stock prices (from $i_3$). The microservices will transform obtained data to a universal format and publish the data to the corresponding topic of the second layer ($s_{1,1}$ to $t_{2,1}$, $s_{1,2}$ to $t_{2,2}$, and $s_{1,3}$ to $t_{2,3}$).

The third layer (Splitter) will have three microservices subscribed to each individual topic in the second layer. Two microservices, $s_{3,1}$ and $s_{3,2}$ (listening to the news topics $t_{2,1}$ and $t_{2,2}$) will keep only the articles related to healthcare and publish them to topic $t_{4,1}$ of the fourth layer. The third microservice $s_{3,3}$ (listening to stock prices in $t_{2,3}$) will publish stock prices of all healthcare companies to topic $t_{4,2}$. The remaining news articles and stock prices will be discarded.

Given that we have two models, $m_1$ and $m_2$, we will have them implemented in two corresponding microservices of the seventh layer. They will listen to topics $t_{6,1}$ and $t_{6,2}$ of the sixth layer, respectively, waiting for the data required for analysis.

Suppose that an analyst decided that $m_1$ requires data on Acme stock prices as well as healthcare-related news articles. Thus, in the fifth layer (Aggregator), we will build a microservice $s_{5,1}$ aggregating the data from topics $t_{4,1}$ and $t_{4,2}$, and then publishing these data to $t_{6,1}$.

Suppose further that an analyst determined that $m_2$ requires data on prices of all healthcare companies and healthcare-related news articles. Then another microservice $s_{5,2}$ in layer 5 will aggregate the data from $t_{4,1}$ and $t_{4,2}$, and then publish it to $t_{6,2}$. We are now ready to test and ship v.1 of our solution.

Note that $s_{5,1}$ filters out all the messages from $t_{4,2}$ that are not related to Acme. Instead, we could have created a new topic in layer 4, to which $s_{3,3}$ would publish only Acme prices. This would decrease the amount of computing. However, as the number of models and companies increases, it may cause excessive fragmentation of topics in layer 4 and make logic of $s_{3,3}$ unmaintainable.

*2) Version 2:* After some period of operation, we received a new requirements: $m_2$ now needs global economic news to better estimate healthcare sector stock portfolio.

To implement this requirement, we will create a new topic $t_{4,3}$ in the fourth layer. We will alter the logic of $s_{3,1}$ and $s_{3,2}$, asking them to publish global economic news to $t_{4,3}$. We will also ask $s_{5,2}$ to start listening to $t_{4,3}$. Note that the rest of the microservices and components remain intact. We are now ready to test and ship v.2 of our solution.

Note that we had to alter four microservices ($s_{3,1}$, $s_{3,2}$, $s_{5,2}$, and $m_2$) and to create one topic $t_{4,3}$; the remaining 17 microservices and topics remained intact. Thus, we can focus testing efforts only on the altered microservices and their downstream counterparts (whereas in the case of monolith application we would have to perform regression testing of the complete solution), hence the reduction of testing efforts and decrease of testing time.

*3) Version 3:* Later on we elicited a new requirement: the analyst now needs to estimate the price of company Globex, operating in a technology sector. To do that, the new model $m_3$ will require global news and stock prices of Globex. To achieve this, we will create a new topic $t_{4,4}$ in the fourth layer and will ask $s_{3,3}$ to publish technology stock prices there. We will also create new microservice $s_{5,3}$ that will aggregate required data from $t_{4,3}$ and $t_{4,4}$. Similar to $s_{5,1}$, $s_{5,3}$ will preserve only Globex prices from $t_{4,4}$. It will then publish the data to newly created topic $t_{6,3}$ in the sixth layer, to which a newly created model $m_3$ will listen to. We are now ready to test and ship v.3 of our solution. Similar to v.2, we focus testing efforts on a subset of created and altered microservices and their downstream counterparts, leading to reduction of testing efforts and shortening of testing phase.

## IV. RELATED LITERATURE

### A. Architecture for Microservices

A significant body of work exists on microservice architecture, see [8] for review. Microservice-based solution easily leverage provisioning of resources as well as elasticity of Cloud infrastructure [11].

Microservice architectures are also cost efficient, in comparison with a monolith one. Villamizar et. al. [20] performed cost comparison study, showing that a microservice solution built with AWS Lambda [17] is 77% cheaper than a monolith application. Balalaie et al. [4] recorded lesson learned on replacing a monolithic backend service by microservices, concluding that microservice architecture reduces the time to deploy new features and improves scalability.

While designing microservice, it is important to define the size of the microservice as it controls the trade-off between scalability and performance [21]. It is also important to keep tightly coupled components in the same microservice to improve performance [21]. To address these, Klock et al. [11] proposed an open-source tool (MicADO) to define scalable microservice, based on a given requirement model and workload. We are leveraging this knowledge in our approach, by implementing individual features of analytic use cases in individual microservices. The objective is to decentralise the loosely coupled use cases and thus reduce the I/O latency of interactions among microservices.

*B. Architecture for Streaming Data*

With the availability of reliable fault tolerant tools and infrastructure, enterprises are building streaming applications to act on real-time data (at least 13 frameworks exist for processing streams [3]). While classic software architecture design principles still apply to analytic applications processing streaming data, the data have direct impact on architectural decision making and quality attributes [15].

Esposito et al. [7] proposed a publish-subscribe-based architecture for streaming data analytics. The authors consider having a single service layer (SL) between a publisher and a subscriber. The key difference between this and our architecture is that we set up multiple layers (instead of a single layer) to funnel out specific data to a proper microservice.

Xhafa et al. [22] created an architecture tailored for Internet of Things based (IoT) systems. It senses, extracts, filters, formats and outputs the data in one single process. This type of architecture is less scalable than microservice-based one and is challenging to implement and maintain. Hromic et al. [10] propose an OpenIoT middleware [19] based approach to collect IoT data and push it to a cluster in a Cloud to apply real-time analysis. This is a use case specific architecture and is not applicable as a general software architecture for streaming data analysis. Moreover, this architecture does data transformation, extraction, and analysis in the same service, making it challenging to alter the code.

Complex event processing based architecture in [5], similarly to our approach, considers "push-based architecture (publish-subscribe)" instead of "pull-based architecture (client/server)". The authors proved that push based architecture is superior to pull based one in terms of accuracy and time. However, [5] is use case centric and has only one service layer. In contrast, our approach can handle multiple use cases within the same architecture and can scale up or down as required.

Nakamoto et al. [16] proposed another IoT-based data centric software architecture. The authors focus on automotive industry. Their approach is similar to ours, as they have data stream management system for all sensors, so that downstream application can get relevant data as required. However, this approach is also based on a single layer (designed for a specific use case), while ours is multi-layered and general.

Simmhan et al. [18] proposed a data-driven analytics platform to meet dynamic demand optimisation of smart electrical grids. The objective is to detect the supply-demand mismatch and correct it on the fly. The idea is to ingest the grid data to a secured repository for researchers and apply pre-trained scalable machine learning model. We cannot compare our architecture with theirs as the details were not provided in [18], but their solution is tailored for a particular use case.

Crowdpulse [14] is a framework doing real time semantic analysis on text of social streams. The authors emphasised semantic analysis of streaming data rather than flexible architecture of streaming application.

To the best of our knowledge, at present, there exists no literature on a generic framework that can handle different use cases for streaming analytics, hence our focus on closing this gap.

V. CONCLUSION AND FUTURE WORK

In this paper we presented a 7-layered architecture for analysis of streaming data and compared it with a 1-layered based architecture (as it is the most prevalent solution today).

The first six layers of the architecture perform data extraction, transformation, filtering, and aggregation. The last, seventh, layer carries analytic models.

Odd layers contain microservices that communicate with their upstream and downstream counterparts asynchronously via topics (hosted by publish-subscribe software) in the even layers. Microservices in a given layer are independent of each other.

This setup ensures low coupling and high cohesion of the solution, leading to its increased scalability and maintainability. However, such setup may cause data duplication, which may lead to space and computing overhead. We perform formal worst-case scenario analysis and derive simple and tractable formulas that show when the 7-layered architecture is more applicable to a business use case than the 1-layered one (from the computational perspective).

Based on our practical experience with financial and e-commerce applications, the 7-layered architecture would be beneficial to a large number of business use cases. Thus, the architecture may aid practitioners, as it allows them to improve their analytic solutions. This work is also a building block for a general architecture for processing streaming data, therefore, it is of interest to academics.

In the future, we plan to explore applicability of this architecture to domains outside of streaming analytics realm.

ACKNOWLEDGMENT

This research is funded in part by NSERC Discovery Grant No. RGPIN-2015-06075.

## REFERENCES

[1] Streaming Analytics Market by Type (Solution & Services), Applications (Fraud Detection, Sales & Marketing Management, Predictive Asset Maintenance, Risk Management, Network Management, Location Intelligence, & Operations Management), Vertical, Regions - Global Forecast to 2021. Technical Report TC3451, Markets and Markets, 2016.

[2] Breadcrumbs Sentry, 2017. https://docs.sentry.io/learn/breadcrumbs/.

[3] M. Babazadeh and C. Pautasso. The Stream Software Connector Design Space: Frameworks and Languages for Distributed Stream Processing. In *2014 IEEE/IFIP Conf. on Softw. Architecture*, pages 1–10, Apr. 2014.

[4] A. Balalaie, A. Heydarnoori, and P. Jamshidi. Microservices Architecture Enables DevOps: Migration to a Cloud-Native Architecture. *IEEE Software*, 33(3):42–52, May 2016.

[5] R. Bhargavi and V. Vaidehi. Semantic intrusion detection with multisensor data fusion using complex event processing. *Sadhana*, 38(2):169–185, Apr. 2013.

[6] K. Birman and T. Joseph. Exploiting Virtual Synchrony in Distributed Systems. In *Proceedings of the Eleventh ACM Symposium on Operating Systems Principles*, SOSP '87, pages 123–138. ACM, 1987.

[7] C. Esposito, M. Ficco, F. Palmieri, and A. Castiglione. A knowledge-based platform for Big Data analytics based on publish/subscribe services and stream processing. *Knowledge-Based Systems*, 79(Supplement C):3–17, May 2015.

[8] P. D. Francesco, I. Malavolta, and P. Lago. Research on Architecting Microservices: Trends, Focus, and Potential for Industrial Adoption. In *2017 IEEE Int. Conf. on Softw. Architecture*, pages 21–30, 2017.

[9] S. Hoque and A. Miranskyy. Architecture for Analysis of Streaming Data: Supplementary Material, 2017. http://scs.ryerson.ca/~avm/dat/sup.pdf.

[10] H. Hromic, D. L. Phuoc, M. Serrano, A. Antonic, I. P. Zarko, C. Hayes, and S. Decker. Real time analysis of sensor data for the Internet of Things by means of clustering and event processing. In *2015 IEEE International Conference on Communications (ICC)*, pages 685–691, June 2015.

[11] S. Klock, J. M. E. M. V. D. Werf, J. P. Guelen, and S. Jansen. Workload-Based Clustering of Coherent Feature Sets in Microservice Architectures. In *2017 IEEE International Conference on Software Architecture (ICSA)*, pages 11–20, Apr. 2017.

[12] V. Lemieux, editor. *Financial Analysis and Risk Management: Data Governance, Analytics and Life Cycle Management*. Springer, 2013.

[13] J. Lewis and M. Fowler. Microservices, 2014. https://martinfowler.com/articles/microservices.html.

[14] C. Musto, G. Semeraro, P. Lops, and M. d. Gemmis. CrowdPulse. *Inf. Syst.*, 54(C):127–146, Dec. 2015.

[15] M. Naab, S. Braun, T. Lenhart, S. Hess, A. Eitel, D. Magin, R. Carbon, and F. Kiefer. Why Data Needs more Attention in Architecture Design - Experiences from Prototyping a Large-Scale Mobile App Ecosystem. In *12th Working IEEE/IFIP Conf. on Softw. Arch.*, pages 75–84, 2015.

[16] Y. Nakamoto, A. Yamaguchi, K. Sato, S. Honda, and H. Takada. Toward Data-Centric Software Architecture for Automotive Systems - Embedded Data Stream Processing Approach. In *2014 IEEE 11th Intl Conf on Ubiquitous Intelligence and Computing and 2014 IEEE 11th Intl Conf on Autonomic and Trusted Computing and 2014 IEEE 14th Intl Conf on Scalable Computing and Communications and Its Associated Workshops*, pages 586–589, Dec. 2014.

[17] P. Sbarski. *Serverless Architectures on AWS: With examples using AWS Lambda*. Manning Publications, Shelter Island, 1 edition, May 2017.

[18] Y. Simmhan, V. Prasanna, S. Aman, A. Kumbhare, R. Liu, S. Stevens, and Q. Zhao. Cloud-Based Software Platform for Big Data Analytics in Smart Grids. *Computing in Science and Engg.*, 15(4):38–47, July 2013.

[19] J. Soldatos, N. Kefalakis, M. Hauswirth, M. Serrano, J.-P. Calbimonte, M. Riahi, K. Aberer, P. P. Jayaraman, A. Zaslavsky, I. P. Zarko, L. Skorin-Kapov, and R. Herzog. OpenIoT: Open Source Internet-of-Things in the Cloud. In *Interoperability and Open-Source Solutions for the Internet of Things*, Lecture Notes in Computer Science, pages 13–25. Springer, Cham, 2015.

[20] M. Villamizar, O. Garcs, L. Ochoa, H. Castro, L. Salamanca, M. Verano, R. Casallas, S. Gil, C. Valencia, A. Zambrano, and M. Lang. Infrastructure Cost Comparison of Running Web Applications in the Cloud Using AWS Lambda and Monolithic and Microservice Architectures. In *2016 16th IEEE/ACM International Symposium on Cluster, Cloud and Grid Computing (CCGrid)*, pages 179–182, May 2016.

[21] E. Wolff. *Microservices: Flexible Software Architecture*. Addison-Wesley Professional, Oct. 2016.

[22] F. Xhafa, V. Naranjo, S. Caball, and L. Barolli. A Software Chain Approach to Big Data Stream Processing and Analytics. In *9th Int. Conf. on Complex, Intelligent, and Softw, Intensive Sys.*, pages 179–186, 2015.

[23] M. Yener and A. Theedom. *Professional Java EE Design Patterns*. John Wiley & Sons, Jan. 2015.